# Metamaterial all-optical switching based on resonance mode coupling in dielectric meta-atoms

Xiaoming Liu, Ji Zhou*
State Kay Lab of New Ceramics and Fine Processing, School of Materials Science and Engineering, Tsinghua University, Beijing, 100084, People's Republic of China
Natalia Litchinitser, Jingbo Sun
Electrical Engineering Department, University at Buffalo, The State University of New York, Buffalo, New York 14260, United States

**Abstract:** All-optical switching based on optical nonlinearity must undergo complex processes of light-mater interaction in atom and electron scale, so a relative high power and long response time is required, that construct main bottlenecks in all-optical technology. In this paper, we propose a new mechanism for building all-optical switching based on a change resonance modes of meta-atoms in metamaterials, not involved optical nonlinearity of materials. A metamaterial composed of cubic dielectric "meta-atoms" periodically embedded in a background matrix is designed and constructed. All-optical switch functionality by modulating the resonance modes in "meta-atoms" utilizing a vertical direction switch light whose power is the same value with that of the signal light is demonstrated in microwave range theoretically and experimentally. This results represent a significant step towards obtaining all-optical switching devices with extremely low threshold power and high switching speed.

———————————————————————————————————————

*Email: zhouji@tsinghua.edu.cn

All-optical switch as a fundamental building block of all-optical digital signal processing system was well studied for decades. Almost all of the previous efforts for this purpose were based on the mechanism of optical nonlinearity, in which the optical properties of the materials was tuned by a pump beam which was strong enough to change materials structure in electronic, atom, molecular or lattice scale. Unfortunately, for most of stable materials, the optical nonlinearity require an really high power (much higher than the signal) and a relative long response time, that construct main bottlenecks in the development of all-optical technology (*1-5*). In this letter, we proposed a nonlinearity-free mechanism of all-optical switching based on a metamaterial route, in which the optical modulation was derived from a change resonance modes, a basic character of meta-atoms in metamaterials, not that of the structure of materials. Therfore, a very low power of light (comparable with signal beam) and extremely short response time (periods of the light wave) are sufficient to generate the all-optical switching process.

Metamaterials are artificial materials engineered to have properties that may not be found in nature, which derived from heir exactingly-designed structures (meta-atoms, metamolecues, meta-crystals) instead of their composition. Unusual electromagnetic properties such as negative refractive, near-field focusing, subwavelength imaging were obtained in metamaterials (*6-12*). Recently, researchers extended the concept metamaterial to conventional material domain, for example, they constructed specialized metamaterials to maintain a nearly constant stiffness per unit mass density even at ultralow density (*13*), to perform mathematical operations (*14*), and to gain high optical nonlinear response (*15*). Optical switching based on metamaterials were widely used to manipulate the propagation of electromagnetic waves (*16-37*). However, they still suffered from high power or long response time. Zhang et al. reported a new approach for light-with-light controlling without nonlinearity by interference in metamaterial (*38*). However, the phenomenon require a coherent controlling beams, that limited the application in devices.

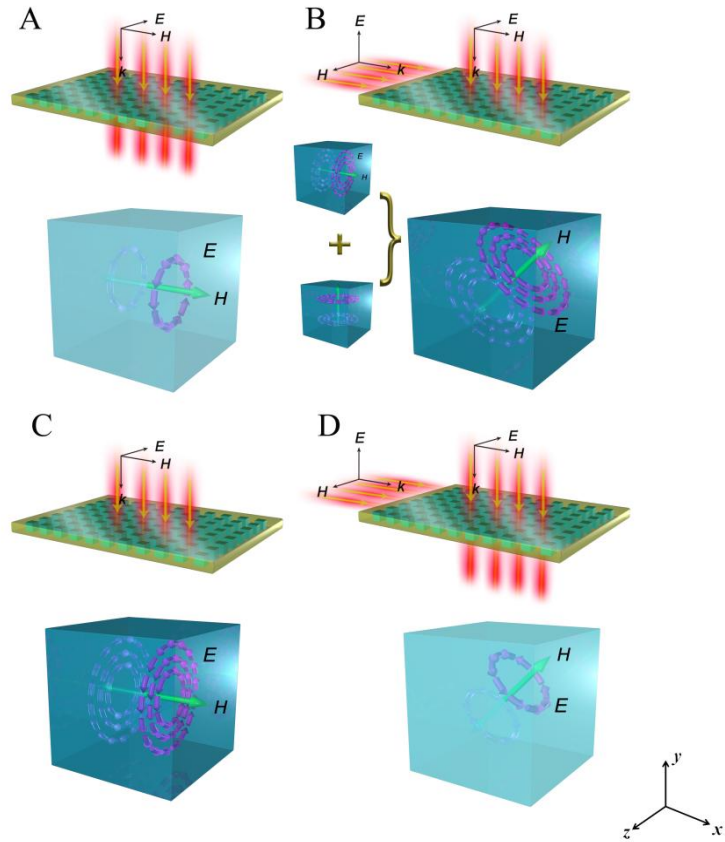

**Fig. 1.** Different resonant modes in dielectric cubes. (A) The meta-switch illuminated by a plane electromagnetic wave at 9.92 GHz with weak resonanc. (B) The meta-switch illuminated by both two waves at 9.92 GHz, a new magnetic resonance mode was induced at lower frequency leading to the switching functionality. (C) The meta-switch illuminated by a plane electromagnetic wave at 9.98 GHz with strong resonanc. (D) The meta-switch illuminated by both two waves at 9.98 GHz with weak resonance.

The mechanism we proposed to generate all-optical switching is based on the modulation of resonance modes of meta-atoms in a dielectric metamaterials. For a dielectric metamaterial with a dielectric cube as meta-atom, different resonant modes was shown in Fig. 1. A meta-atom with a permittivity $\varepsilon_2$ and a permeability $\mu_2$ embedded in a background matrix with the permittivity $\varepsilon_1$ and permeability $\mu_1$ was illuminated by a plane electromagnetic wave propagating in *y* direction with magnetic field polarized in *x* direction (Fig. 1A), so that for the incident field

$$H_{xi} = |H_i| e^{-jky\sqrt{\varepsilon_1 \mu_1}} \quad (1)$$

The scattered field of the cube at $(x_0, y_0, z_0)$ can be put in the form

$$H_{xs} = [A(l)]^3 \left[ H(y_0) \frac{\mu_p - \mu_1}{\mu_p + 2\mu_1} \left( \frac{\partial^2}{\partial x^2} + k^2 \mu_1 \varepsilon_1 \right) - jk\varepsilon_1 E(y_0) \frac{\varepsilon_p - \varepsilon_1}{\varepsilon_p + 2\varepsilon_1} \frac{\partial}{\partial y} \right] \frac{e^{-jkr_0\sqrt{\varepsilon_1\mu_1}}}{r_0} \quad (2)$$

$$r_0 = \sqrt{(x-x_0)^2 + (y-y_0)^2 + (z-z_0)^2} \quad (3)$$

$$\frac{\varepsilon_p}{\varepsilon_2} = \frac{\mu_p}{\mu_2} = \frac{2(\sin\theta - \theta\cos\theta)}{(\theta^2 - 1)\sin\theta + \theta\cos\theta} \quad (4)$$

$$\theta = kA(l)\sqrt{\varepsilon_2\mu_2} = 2\pi f A(l)\sqrt{\varepsilon_2\mu_2}/c \quad (5)$$

$c$ is the speed of light, $k$ is the wavenumber in the host medium, $l$ is the size parameter in the resonance direction, $A(l)$ is a monotonically increasing function of $l$. The field exciting the cube at $(x_0, y_0, z_0)$ is given by the sum of the incident field and the scattered field by other cubes.

$$H_x = H_{xi} + \sum H_{xs} \quad (6)$$

The first resonance is magnetic with its resonance direction along x axis, so $l = a$ mm, According to Eq (5), the frequency of the first resonance is

$$f_1 = \frac{\theta_1 c}{2\pi A(l)\sqrt{\varepsilon_2\mu_2}} = \frac{\theta_1 c}{2\pi A(a)\sqrt{\varepsilon_2\mu_2}} \quad (7)$$

$\theta_1$ is a constant. The switch plane electromagnetic wave propagated in x direction with magnetic field polarized in z direction illustrated in Fig. 1B. The field exciting the cube at $(x_0, y_0, z_0)$ is

$$H_z = H_{zi} + \sum H_{zs} \quad (8)$$

When the dielectric metamaterial was illuminated by both the signal and switch waves, the field can be described as

$$H = H_x + H_z = H_{xi} + H_{zi} + \sum H_{xs} + \sum H_{zs} \quad (9)$$

The magnetic resonance is along diagonal direction between x and z axes, so $l = \sqrt{2}a$. According to Eq (5), the frequency of the first resonance is

$$f_1' = \frac{\theta_1 c}{2\pi A(l)\sqrt{\varepsilon_2\mu_2}} = \frac{\theta_1 c}{2\pi A(\sqrt{2}a)\sqrt{\varepsilon_2\mu_2}} \quad (10)$$

$A(l)$ is a monotonically increasing function of $l$, $A(\sqrt{2}a) > A(a)$, $f_1' < f_1$. Therefore, the magnetic resonance frequency will shift to lower frequency because of the switch wave, which leads to the switching functionality.

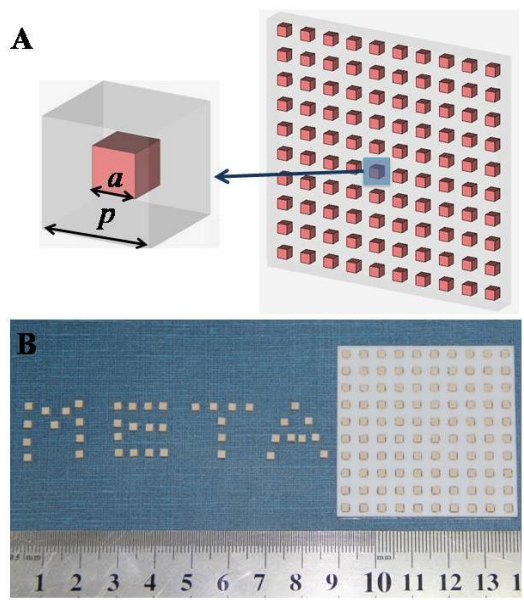

**Figure. 2.** The schematic of the all-optical dielectric meta-switch. (A) The designed meta-switch and its single cubic unit cell. (B) The experimental sample of the meta-switch composed of one hundred CaTiO$_3$ cubes embedded in background ABS matrix

The schematic of the designed all-optical dielectric meta-switch and its single cubic unit cell are demonstrated in Fig. 2A. The switch consists of a matrix of acrylonitrile butadiene styrene (ABS) plastic ($\varepsilon_1$=2.67, tan$\delta_1$=0.006), with dielectric cubes regularly embedded within it. The cubes are made of microwave ceramics (CaTiO$_3$-0.5%wt ZrO$_2$), which possess high permittivity ($\varepsilon_2$=168.5) and low loss (tan$\delta_2$=0.001). $a$ is the side length of the cubes, $c$ is the lattice constant which is much smaller than the wavelength of the incident electromagnetic wave. By explicit simulation, we achieved dimensions of the unit cell, in millimeters, of: $a$ = 2, $p$ = 5, which make the magnetic resonance of the dielectric metamaterial occur at 9.98 GHz. We could easily scale the design to lower RF and higher optical frequencies by changing the permittivity and size of the dielectric particles.

Computer simulations were performed using the radio frequency model of Comsol Multiphysics 4.4. First, we investigated the S-parameter of transmission [$S_{21}(\omega)$] of the dielectric meta-switch illuminated only by signal wave as shown in Fig. 1A、C and examined the behavior of the magnetic field in the dielectric cubes. Fig. 3A (red curve) indicate that we achieve a resonance transmission spectrum with the minimum at 9.98 GHz resulting from the magnetic resonance (Fig. 4A,D) along $x$ direction induced by signal wave. Then we investigated the S-parameter of transmission [$S_{21'}(\omega)$] of the dielectric metamaterial illuminated by switch wave in the absence of signal wave, the simulated result was shown in Fig. 3B (red curve), which revealed that the intensity of the wave received by port 2 is extremely small because the switch wave propagate along $x$ axis. However, the behavior of the magnetic field of the dielectric cube at 9.98 GHz in Fig. 4B,E demonstrated that a strong magnetic resonance along $z$ direction was induced by the switch wave. Finally, when the dielectric meta-switch was illuminated by both the signal and switch electromagnetic waves as illustrated in Fig. 1B、D, the magnetic resonances of the dielectric metamaterial can be induced along *x* and *z* direction respectively by the signal and switch wave. The coupling effects of the magnetic dipoles derived by the two waves lead to a new magnetic resonance mode along the cube's diagonal line, which resulted in a transmission minimum at 9.92GHz. Fig. 3C (red curve) and Fig. 4C,F depicted the S-parameter of transmission $[S_{2(1+1')}(\omega)]$ of the dielectric metamaterial by two waves and the behavior of the simulated magnetic field of the new resonance mode at 9.92GHz.

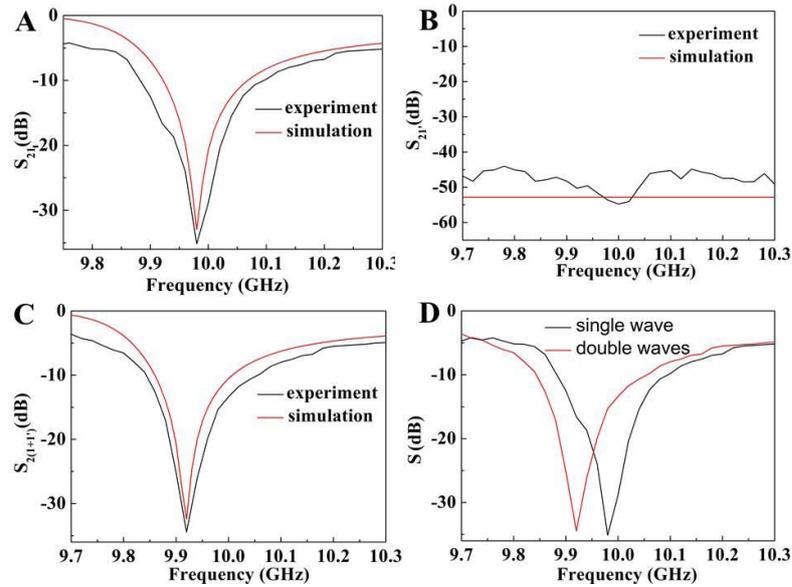

**Fig. 3.** Simulated (red line) and experimental (black line) S-parameter of transmission of the dielectric meta-switch illuminated by (A) signal wave, (B) switch wave and (C) both the two waves. (D) switch-on and switch-off state at 9.92 and 9.98 GHz.

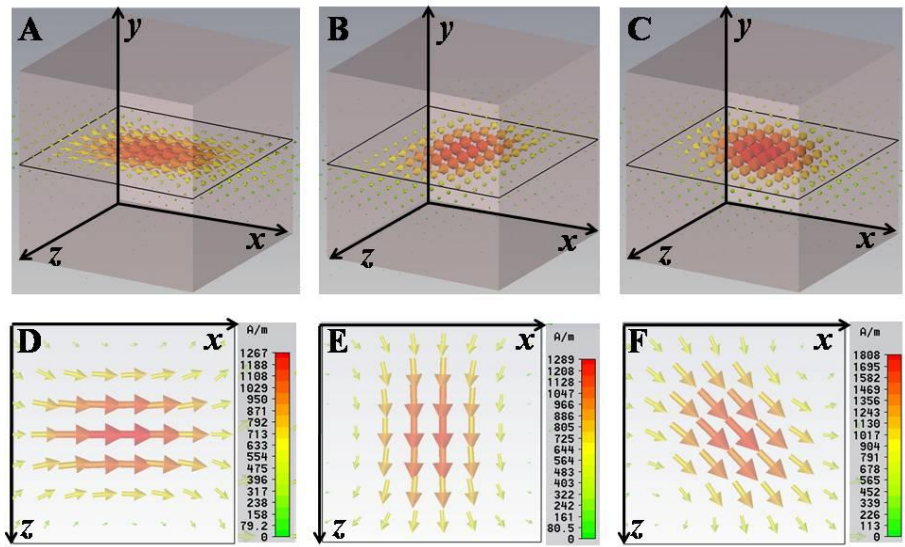

**Fig. 4.** The simulated behavior of the magnetic field in the dielectric cube at resonance frequency. (A)/(D) Perspective/top view of the magnetic field distribution at 9.98 GHz induced by signal wave. (B)/(E) Perspective/top view of the magnetic field distribution at 9.98 GHz induced by switch wave. (C)/(F) Perspective/top view of the magnetic field distribution at 9.92 GHz induced by both the two waves.

A dielectric ceramic material CaTiO$_3$ was synthesized by solid state reaction method and cut into cubes whose size parameters and permittivity are the same with the simulated values. The all-optical meta-switch sample was composed of one hundred CaTiO$_3$ cubes embedded in background ABS matrix as shown in Fig. 2B. The microwave S-parameter of transmission of the dielectric metamaterial was measured by two sources network analyzer Agilent N5242A PNA-X using a linearly polarized horn antennas. The power of the signal and switch wave was set to be 1mW as we simulated. When the sample was illuminated by signal wave, switch wave and both the two waves, a good agreement between experimental (black curve) and simulated (red curve) S-parameter of transmission was achieved as illustrated in Fig.3A, B and C (black line) which indicated that we successfully prepared an all-optical dielectric meta-switch as we had designed by simulation. The switch-on and switch-off state at 9.92 and 9.98 GHz were demonstrated clearly in Fig.3D. There was a high transmission corresponding to the switch-on state at 9.92 GHz when the dielectric meta-switch was illuminated just by signal wave (Fig.1A). While by both the signal and switch waves, we can achieve a relatively low transmission at the same frequency considered as the switch-off state (Fig. 1B). In a similar way, There was a low transmission corresponding to the switch-off state at 9.98 GHz when the dielectric meta-switch was illuminated just by signal wave (Fig.1C). While by both the signal and switch waves, we can achieve a relatively high transmission at the same frequency considered as the switch-on state (Fig. 1D).

In conclusion, we proposed a new mechanism all-optical switching without optical nonlinearity process. The switching is derived from the couple of resonance modes in the meta-atoms of metamaterials. An all- optical modulation in microwave range was demonstrated numerically and experimentally in a dielectric metamaterial composed of dielectric cubes as the meta-atoms, in which the couple occurs between magnetic resonances of meta-atom induced along *x* and *z* direction respectively by the signal and switch wave. The coupling of the magnetic dipoles derived by the two waves lead to a new magnetic resonance mode along the cube's diagonal line at lower frequency. There was a high transmission corresponding to the switch-on state at 9.92GHz when the dielectric meta-switch was illuminated just by signal wave. While by both the signal and switch waves, we can achieve a relatively low transmission at the same frequency considered as the switch-off state. This all-optical switch mechanism based on the modulation of resonance mode of meta-atoms should be applicable to much high frequency till to visible range, representing a significant step towards obtaining all-optical switching devices with extremely low threshold power and high switching speed.

**Acknowledgements**

This work was supported by the National Natural Science Foundation of China under Grant Nos. 51032003, 11274198, 51221291 and 61275176, and National High Technology Research and Development Program of China under Grant No. 2012AA030403.